\documentclass{ws-procs975x65}

\begin{document}

\title{THIRTY YEARS OF STUDIES OF INTEGRABLE REDUCTIONS\\ OF EINSTEIN'S FIELD EQUATIONS}

\author{G.A. ALEKSEEV}

\address{Steklov Mathematical Institute RAS,\\
 Gubkina 8, 119991, Moscow Russia\\
E-mail: G.A.Alekseev@mi.ras.ru}

\begin{abstract}

More than thirty years passed since the first discoveries of various aspects of integrability of the symmetry reduced vacuum Einstein equations and electrovacuum Einstein - Maxwell equations were made and gave rise to constructions of powerful solution generating methods for these equations. In the subsequent papers, the inverse scattering approach and soliton generating techniques, B\"acklund and symmetry transformations, formulations of auxiliary Riemann-Hilbert or homogeneous Hilbert problems and various linear integral equation methods have been developed in detail and  found different interesting applications.
Recently many efforts of different authors were aimed at finding of  generalizations of these 
solution generating methods to various (symmetry reduced) gravity, string gravity and 
supergravity models in four and higher dimensions. However, in some cases it occurred that 
even after the integrability of a system was evidenced, some difficulties arise which do not 
allow the authors to develop some effective methods for constructing of solutions.  
The present survey includes some remarks concerning the history of discoveries of some of 
the well known solution generating methods for these equations, brief descriptions of various 
approaches and their scopes as well as some comments concerning the possible difficulties 
of generalizations of various approaches to more complicate gravity models and possible 
ways for avoiding these difficulties.
\end{abstract}


\bodymatter

\section*{Discovery of Integrability of Einstein's field Equations
\footnote{${}^{\scriptscriptstyle )}$ It is necessary to clarify that this integrability was 
discovered only for the "symmetry reduced" Einstein's field equations, what means that these 
equations are considered in $D=4$ space-times with two commuting isometries (or, more 
generally, in space-times of $D\ge 4$ dimensions with $D-2$ commuting isometries) such 
that one of them is time-like (the "elliptic" reduction, e.g., stationary axisymmetric fields) or 
all isometries are space-like  ("hyperbolic" reduction, e.g., plane, cylindrical or other types of 
waves,  cosmological models.) In this case, it is assumed also that all field components and 
their potentials depend only on two of $D$ space-time coordinates.}${}^{\scriptscriptstyle )}$}

In mathematical physics, the period since the middle of 60th years of the previous century was marked by a wonderful discovery of existence of very interesting class of nonlinear partial differential equations which were called completely integrable and which admit various powerful solution generating methods for explicit construction  of infinite hierarchies of solutions with arbitrary large number of parameters (e.g., such as multi-soliton and finite-gap solutions), constructing some nonlinear superpositions of fields (solitons interacting with arbitrary backgrounds), solving  various initial and boundary value problems. It was the more important because two-dimensional reductions of many fundamental equations from different areas of mathematical and theoretical physics were found to belong to this class.\footnote{${}^{\scriptscriptstyle )}$ For the readers who are not closely familiar with this context, it seems useful to clarify, avoiding rigorous definitions, that the notion of (complete) integrability for some nonlinear system of partial differential equations does not mean necessarily  that the solution can be found explicitly by a direct integration.  In general, the integrability of a nonlinear system of partial differential equations means that it possess very rich internal structure such that it admits an infinite set of conservation laws, infinite dimensional algebra of internal symmetries, a representation in terms of equivalent linear (spectral) problem (e.g., Lax-pairs, AKNS or Zakharov - Shabat U-V linear systems), various "dressing" procedures, such as, e.g.,  soliton or B\"acklund transformations acting on the whole space of solutions, the so called "prolongation structures" as well as many other features. Moreover, in the infinite-dimensional space of solutions of an integrable system, it occurs possible to find such transformation of "coordinates" (i.e. of the functional parameters which characterize every solution) that in these new "coordinates" the transformed nonlinear equations can be trivially solved. Thus, the original nonlinear problem transforms into the problem of constructing such "coordinate transformation" -- the so called "direct" problem (to find such "coordinates" for every given solution) and the "inverse" problem (to find the solution for any given values of these new "coordinates"). For various known integrable systems the functional parameters which can be chosen as such useful new "coordinates" can possess different character: these are, for example, the scattering data as functions of the spectral parameter in the well known inverse scattering transform (called also Inverse Scattering Method or simply ISM), the Riemann - Hilbert data -- a set of arbitrarily chosen functions on the contour on the plane of auxiliary free complex ("spectral") parameter in various formulations of equivalent Riemann - Hilbert problems, the monodromy data for the fundamental solution of associated linear system as functions of the "spectral" parameter in the monodromy transform approach, etc. It is remarkable, that in all cases, this allows to reduce the problem of solution of the original nonlinear equations to solution of the direct and inverse problems of these transforms which imply to solve the linear equations only: for solution of the direct problem one has to solve an overdetermined  linear system of differential equations with a free  "spectral" parameter, while the solution of the inverse problem can be reduced to solution of some linear integral equations (such as Gelfand - Levitan - Marchenko equation in the case of traditional formulation of ISM, or some systems of linear singular integral equations or (quasi-) Fredholm linear integral equations in other cases.
These "coordinate transforms" extend considerably a variety of mathematical tools which can be used for solution of given nonlinear system and allow, in particular, to develop different effective methods for constructing  infinite hierarchies of solutions, to perform asymptotical analysis of solutions and to construct solutions for various initial and boundary value problems.} ${}^{\scriptscriptstyle )}$

\subsubsection*{Integrability conjectures}
Further progress in the theory of completely integrable nonlinear equations and of its applications at the beginning of 70th years of 20th century gave rise to obvious hopes and resonable expectations that the Einstein's field equations at least for the  simplest cases of pure vacuum or electrovacuum space-times with Abelian two - dimensional isometry group (stationary axisymmetric fields, plane waves, etc) can happen to be also integrable. Thus in 1972 Geroch \cite{Geroch:1972} conjectured  that vacuum Einstein equations with two commuting Killing vector fields admit an infinite-dimensionsl group of internal symmetries
which allows to obtain any solution starting from Minkowski space-time.
Later, in a series of papers Kinnersley \cite{Kinnersley:1977} and then Kinnersley and Chitre \cite{KCII:1977}${}^{-}$\cite{ KCIV:1978b} studied the internal symmetries of stationary axisymmetric Einstein - Maxwell equations. These authors descovered the existence of an infinite dimensional algebra of internal symmetries of these equations and constructed its representation in terms of infinite hierarchies of potentials which characterize every solution. For vacuum case, they had found in a closed form some elements of the corresponding infinite dimensional group of symmetry transformations. It is worth to mention also a wonderful finding in the last mentioned paper  of Kinnersley and Chitre \cite{ KCIV:1978b} where these authors observed (a) that the infinite hierarchy of potentials which correspond to a given stationary axisymmetric solution of vacuum Einstein equations admits a $2\times 2$ - matrix generating function, depending on Weyl coordinates and a free complex parameter and (b) that this generating function should satisfy some system of linear partial differential equations.
However, in the papers \cite{KCII:1977}${}^{-}$\cite{ KCIV:1978b} the authors concentrated on the group-theoretical approach and the question whether the linear system for generating function found in \cite{KCIV:1978b} can play the role of associated spectral problem did not arose in these papers.

In less than two weeks after the paper \cite{ KCIV:1978b} of Kinnersley and Chitre was submitted for publication, another  paper of a different author was submitted to another journal. It was the paper \cite{Maison:1978} of  Maison with the title \emph{"Are the stationary, axially symmetric Einstein equations completely integrable?"}, where the author, using completely different approach, constructed for vacuum equations some "linear eigenvalue problem in the spirit of Lax" and noticed that this "nourish some hope that a method similar to the inverse-scattering method may be developed" and that this would allow to reduce the solution of these nonlinear equations to a sequence of linear problems.

The papers mentioned above concluded the period of integrability conjectures, because very soon after these papers were submitted and before their publication, Belinski and Zakharov submitted a paper where the integrability of vacuum Einstein equations with two commuting isometries was proved and began "to work".

\subsubsection*{Discovery of integrability of vacuum Einstein equations}
In their well known paper, Belinski and Zakharov \cite{Belinski-Zakharov:1978}
constructed a new overdetermined linear system with a free complex ("spectral") parameter whose integrability conditions are equivalent to the symmetry reduced vacuum Einstein equations. The conditions of reality and symmetry of metric coefficients were reformulated as the explicit constraints ("reduction conditions") on the space of solutions of this associated linear system. Thus the original nonlinear equations werereformulated in terms of \textit{equivalent} spectral problem. Using this spectral problem, these authors\\[-3ex]

\leftskip5ex
\noindent\llap{(a)\hskip1ex}\hskip0ex discovered the existence of gravitational solitons and developed a "dressing method" for constructing of N solitons on any chosen background;\\[-3ex]

\leftskip5ex
\noindent\llap{(b)\hskip1ex}\hskip0ex reformulated the spectral problem in terms of a classical matrix Riemann problem, i.e. the problem of finding of two matrix functions of one complex variable which are analytical respectively inside and outside of some closed contour on the complex (spectral) plane and which satisfy given matching conditions on the contour. These matching conditions include some functional parameters ("the contour data") which characterize any particular solution of (symmetry reduced) vacuum Einstein equations;\\[-3ex]

\leftskip5ex
\noindent\llap{(c)\hskip1ex}\hskip0ex they had transformed also the constructed Riemann problem to a system of linear singular integral equations of a Cauchy type such that every particular solution of these integral equations determines some solution of (symmetry reduced) vacuum Einstein equations and vise versa.\\[-2ex]

\leftskip0ex\noindent
Thus, in this paper, the first effective algorithm for constrcuting of infinite hierarchies of exact (N-soliton) solutions on arbitrary chosen background was suggested and the main directions for further developments were discovered and outlined.

\section*{Many "languages" of integrability}
Many different mathematical structures, formalisms and mathematical languages which can be associated with a completely integrable system, may give rise to development of different approaches to its analysis and lead to some new ways for its solution. For vacuum Einstein equations a number of different approaches began to develop then independently by different authors. Some of these approaches were generalized to electrovacuum Einstein - Maxwell euqations which also occur to have integrable structure similar in many aspects to that of vacuum equations. Below we mention only the earliest publications in these directions, concentrating in this section on vacuum and electrovacuum cases only.

\subsubsection*{Inverse scattering method} Belinski and Zakharov  soliton generating algorithm \cite{Belinski-Zakharov:1978}{} provides the explicit expressions for all metric components besides the coefficient in the conformally flat part of the metric. This conformal factor for vacuum solitons was found in \cite{Belinski-Zakharov:1979}{} in an explicit (determinant) form. All other metric components of vacuum 2N-soliton solution\footnote{${}^{\scriptscriptstyle )}$ In Belinski and Zakharov technique, the signature of metrics of soliton solutions generating on the background of the Lorenz signature will be also Lorenz only if the number of solitons is even.}${}^{\scriptscriptstyle )}$ also can be presented in a determinant  form \cite{Alekseev:1981} which includes only three determinants of  $2N\times 2N$ - matrices.

For electrovacuum Einstein-Maxwell equations the Belinski and Zakharov approach did not admit a direct generalization.\footnote{${}^{\scriptscriptstyle )}$ As it was shown in \cite{Alekseev:1983a}{}, the Belinski and Zakharov linear system (considered not for $2\times 2$ real symmetric, but for Hermitian $3\times 3$-matrices) can be used as spectral problem for symmetry reduced Einstein - Maxwell equations, however, an appropriate reduction conditions providing equivalence to Einstein - Maxwell equations have not been found in an appropriate form.}${}^{\scriptscriptstyle )}$ Therefore, to apply the inverse scattering ideas for this case it was necessary to find some new form of this approach. The spectral problem for the symmetry reduced Einstein - Maxwell equations and the corresponding electrovacuum $N$-soliton solutions on arbitrary electrovacuum background were constructed in \cite{Alekseev:1980a}${}^{,}$\cite{Alekseev:1980b}{}. This spectral problem, being restricted to vacuum case and compared with Belinski and Zakharov one, shows essentially different structure. In particular, it does not include differentiation with respect to a spectral parameter and the plane of the spectral parameter introduced in \cite{Belinski-Zakharov:1978} ("$\lambda$-plane") covers twice the spectral plane introduced in \cite{Alekseev:1980a}${}^{,}$\cite{Alekseev:1980b} ("$w$-plane"). The electrovacuum solitons \cite{Alekseev:1980a}${}^{,}$\cite{Alekseev:1980b} necessary have complex (but not real) poles and the vacuum part of these $N$-soliton solutions (generated on a vacuum background) coincide with $2 N$-soliton solutions of Belinski and Zakharov generated on the same vacuum background with $N$ pairs of complex conjugated poles. The electrovacuum generalizations of Belinski and Zakharov solitons with real poles do not arise in this algorithm, however, for appropriate choice of soliton parameters, such solutions arise as trivial analytical continuations of solitons with complex poles in the space of their parameters.\\[-3ex]

\subsubsection*{B\"acklund and symmetry transformations}
A few months after the Belinski and Zakharov paper was submitted, Harrison (who was already acquainted with Belinski and Zakharov approach and cited this in his paper), submitted a paper  \cite{Harrison:1978}{} where he used a general approach of Estabrook and Wahlquist \cite{Estabrook-Wahlquist:1976} based on the construction of the so called ``prolongation structures'' for nonlinear evolution equations. In this paper, Harrison constructed the B\"acklund transformations for vacuum Ernst equations. It is worth to mention here that a nice technical improvement of this formalism based on the use of the so called constant coefficient ideals (``CC-ideals'') of one-forms  was suggested  by Harrison in a later paper \cite{Harrison:1983}{}, where he presented also a generalization of his construction of vacuum B\"acklund transformations to electrovacuum Einstein - Maxwell fields.

Very soon after the paper \cite{Harrison:1978} was submitted, but before its publication, Neugebauer submitted a paper
\cite{Neugebauer:1979} where he also constructed B\"acklund transformations for vacuum Ernst equations, using different approach based on subsequent application of simple non-commuting groups of transformation. In the subsequent papers \cite{Neugebauer:1980a}${}^,$\cite{Neugebauer:1980b}{}, this approach was developed in more details and also gave rise to generation of solutions with arbitrary number of independent parameters.   In particular, for  $N$-fold B\"acklund transformation the generated Ernst potential was expressed in terms of two determinants of $(2 N+1)\times (2 N+1)$ order.{}\footnote{${}^{\scriptscriptstyle )}$It can be shown that these solution generating transformations coincide actually with the soliton generating transformations constructed by Belinski and Zakharov.}${}^{\scriptscriptstyle )}$

In the papers of Julia \cite{Julia:1983, Julia:1985} the infinite dimensional symmetry transformations of Geroch and Kinnersley were recognized as Kac-Moody symmetries. These  papers included also a reach material which provides much wider vision of the subject including the enlargement of the symmetry groups, the structure of generalized $\sigma$-models and supergravities in various dimensions. In the paper of Breitenlohner and Maison \cite{Breitenlohner-Maison:1987} the structure of the corresponding infinite-dimensional Geroch group was described in detail.

\subsubsection*{Integral equation methods for effecting Kinnersley-Chitre transformations}
A few months after the first of the above mentioned papers were published, Hauser and Ernst suggested the integral equation method for exponentiating the Kinnersley and Chitre infinitesimal symmetries to finite solution generating transformations,\footnote{${}^{\scriptscriptstyle )}$ Though this approach  was constructed in essentially different context of Geroch and Kinnesrley and Chitre symmetry transformations, the basic ideas used there for reformulation of the problem in terms of a homogeneous Hilbert problem and corresponding singular integral equations seem to be very close to those used by  Belinski and Zakharov in their constrcution of a matrix Riemann problem and linear integral equations in the framework of their inverse scattering approach.}${}^{\scriptscriptstyle )}$ where, in particular,
\begin{list}{$\diamond$}{\leftmargin4ex}
\item{it was found \cite{Hauser-Ernst:1979a}{} that a matrix generating function (``F-potential'') constructed earlier by Kinnersley and Chitre\cite{KCIV:1978b} for an infinite hierarchy of potentials associated with any particular vacuum solution, should satisfy some $2\times 2$-matrix linear singular integral equation of a Cauchy type. The matrix kernel of this integral equation depends on the F-potential of a chosen ``seed'' solution and on some set of coordinate-independent functional parameters which characterize any symmetry transformation. Given seed solution and given values of these functional parameters, the solution of this matrix integral equation determines the generating function (F-potential) of the Kinnersley-Chitre hierarchy of potentials associated with the transformed solution.}
\item{In the next paper \cite{Hauser-Ernst:1979b}{}, the authors constructed $3\times 3$-matrix generalization of the F-potential which characterizes every stationary axisymmetric electrovacuum solution and found a linear differential equation for this F-potential. In the same paper \cite{Hauser-Ernst:1979b}{}, these authors presented the corresponding electrovacuum $3\times 3$-matrix generalization of the $2\times 2$-matrix linear singular integral equation constructed in \cite{Hauser-Ernst:1979a}{} for pure vacuum case.
Besides that, in this paper the authors suggested another derivation of their integral equation for the F-potential which is not based on infinitesimal symmetries and Kinnersley-Chitre infinite hierarchies of potentials.}
\item{In\cite{Hauser-Ernst:1980a}${}^{,}$\cite{Hauser-Ernst:1980b}{}, the authors presented a more detail development of their linear singular integral equation method and reduce the problem of constructing the symmetry transformations of the so called Geroch group to a classical problem for analytical functions of one complex variable -- the matrix homogeneous Hilbert problem ($2\times 2$-matrix problem for vacuum case or $3\times 3$-matrix problem for electrovacuum case).}
\item{In the subsequent papers \cite{Hauser-Ernst:1980c}${}^{,}${}\cite{Hauser-Ernst:1981}{}, Hauser and Ernst expressed explicitly the functional parameters of the symmetry transformation which maps any given stationary axisymmetric vacuum or electrovacuum  solution into any other one in terms of the values of the Ernst potentials of these seed and transformed solutions on some regular part of the axis of symmetry and proved in this way the Geroch conjecture that any of such solutions can be generated by a symmetry transformation from, e.g., the Minkowski space-time.\footnote{${}^{\scriptscriptstyle )}$ We note here that in the Hauser and Ernst papers \cite{Hauser-Ernst:1980a}${}^{-}$\cite{Hauser-Ernst:1980c}{}, many mathematical details of the construction of the homogeneous Hilbert problem and of the derivation of their integral equation method were elaborated in many interesting and useful details. However, in some other points the authors used (clearly formulated) ``working hypothesis'', which proof was not known to the authors. These concerned, first of all, the zero values of the component indices of the constructed homogeneous Hilbert problem and therefore, the existence of the solution of the derived singular integral equations for any choice of the seed solution and transformation parameters. Another such point concerns the existence of the gauges which minimize the F-potential singularities on the plane of auxiliary complex parameter. It seems that a conjecture of existence of such gauges restricts the class of stationary axisymmetric vacuum and electrovacuum fields by the solutions whose domains include some regular part of the axis of symmetry. This conjecture looks physically very natural for stationary axisymmetric fields, but it is not available for a similar class of fields with two commuting space-like Killing vectors, such as, for example, colliding plane waves or cosmological models in which we have respectively, instead of the axis of symmetry,  the  focusing or initial curvature singularities.}${}^{\scriptscriptstyle )}$}
\item{In \cite{Hauser-Ernst:1979b} it was shown also by concrete examples that if the functional parameters in the kernel of the integral equation are chosen as \textit{rational} functions of a complex parameter, the integral equation can be solved analytically and the corresponding transformation of the seed solution can be found explicitly.}
\end{list}

A few years later, Sibgatullin   \cite{Sibgatullin:1984a}${}^{,}$\cite{Sibgatullin:1984b} suggested a  modification of the Hauser and Ernst integral equation method which facilitated considerably its application to calculation of particular solutions if the ``input data'' for a constructing solution are the values of its Ernst potentials on the axis of symmetry given as \textit{rational} functions of the canonical Weyl coordinate $z$ along the axis. For this it was suggested
\begin{romanlist}
\item{to refuse for simplicity of the idea to have in the kernel of the integral equation an arbitrary seed and to restrict a consideration by the Minkowski space-time as the seed solution and}
\item{to express (using the way identical to that described in Hauser and Ernst papers \cite{Hauser-Ernst:1980c}${}^{,}$\cite{Hauser-Ernst:1981}{}) the Geroch group functional parameters in the kernel of the Hauser and Ernst matrix integral equation in terms of the input data -- the axis values of the Ernst potentials of the transformed solution.}
\end{romanlist}
With such modification, the Hauser and Ernst matrix integral equation (restricted to the Minkowski seed) can be reduced to a scalar integral equation for  one unknown function, supplied with a ``normalization condition'' also imposed on this function.\footnote{${}^{\scriptscriptstyle )}$Derivation of the integral equation given in \cite{Sibgatullin:1984a}${}^{,}$\cite{Sibgatullin:1984b} by Sibgatullin had not been supplied by any further development of the theory of integrability of the Ernst equations and therefore, its applicability also restricted by the same ``working hypothesis'' used by Hauser and Ernst for derivation of their matrix integral equation (see the comments in previous footnote). Moreover, such input data for the solutions as the values of the Ernst potentials on some regular part of the axis of symmetry outside the sources, typically can not be defined in advance, directly from some physically motivated conditions. Besides that, in this case, all solutions are defined only in the domains which must include some regular part of the axis of symmetry and this clearly do not allow to begin with the boundary conditions given anywhere outside the axis.}${}^{\scriptscriptstyle )}$

Choosing some rational values of the Ernst potentials on a regular part of the axis of symmetry, assuming an appropriate rational structure of the desirable solution of the integral equation on the complex plane and using the well known properties of Cauchy principal vale integrals and the theory of residues, the integral equation can be reduced to an algebraic system. It is clear that such algebraic system is much simpler for the case of scalar integral equation derived by Sibgatullin than for the original matrix integral equation of Hauser and Ernst. Accordingly,
in the last two decades, this integral equation was actively exploited by Sibgatullin and his co-authors for calculation of particular examples of asymptotically flat stationary axisymmetric vacuum and electrovacuum solutions with rational axis data.\footnote{${}^{\scriptscriptstyle )}$ It is worthy to note that the soliton generating techniques (for vacuum \cite{Belinski-Zakharov:1978}${}^{,}$\cite{Belinski-Zakharov:1979} as well as for electrovacuum\cite{Alekseev:1980a}${}^{,}$\cite{Alekseev:1980b})  which had been developed much earlier, being applied to a Minkowski seed, also give numerous stationary axisymmetric asymptotically flat solutions with rational axis data. It is not difficult to show that at least in vacuum case, most of the solutions calculated from the integral equations (or even all of them) can be identified as the particular cases of soliton solutions or are the limits of the soliton solutions, as in the case of multiple poles. The situation with electrovacuum solutions may seem to be not so simple because, as it was already mentioned above, a direct application of the electrovacuum soliton generating technique \cite{Alekseev:1980a}${}^{,}$\cite{Alekseev:1980b} do not lead directly to solitons with real poles on arbitrary chosen background (i.e. to electrovacuum generalization of Belinski and Zakharov vacuum solitons with real poles). However, it is obvious, that all such solitons on the Minkowski background can be constructed as analytical continuations of solitons with complex conjugated poles in the space of their (appropriately chosen) parameters or as the limiting cases of these solutions, as in the case of multiple poles. This analytical continuation is completely similar to the well known analytical continuation which connects the ``underextreme'' part of the family of Kerr-Newman solutions representing charged rotating black holes with the ``overextreme'' part of this family representing naked singularities.
However, it was natural to expect that various integral equation methods may find other interesting applications in different studies of solutions of integrable reduction of Einstein and Einstein - Maxwell equations and these expectations were actually confirmed by the subsequent developments.}${}^{\scriptscriptstyle )}$

\subsubsection*{Monodromy transform approach}

\noindent
Similarly to the inverse scattering method, the monodromy transform approach presented in \cite{Alekseev:1985}${}^{,}$ \cite{Alekseev:1986} (for the proofs and outlines see \cite{Alekseev:1988}${}^{-}$\cite{Alekseev:2005a} ) also starts from a formulation of some ``spectral problem'' -- an overdetermined linear system with a complex parameter supplied with some constraints which provide an equivalence of the entire spectral problem to the Einstein's field equations.

The key point of this approach is a definition of the monodromy data as a set of coordinate-independent functions of a complex parameter which  characterize \emph{any} local solution\footnote{${}^{\scriptscriptstyle )}$ The ``local solution'' means that all metric components and field potentials are considered in some neighborhood of a given space-time point where these are analytical functions of coordinates. The ``space of local solutions'' at a given point means a set of all possible local solutions (each with its own local domain) at the same ``reference'' point where all field variables as well as the fundamental solution of the spectral problem are normalized by some standard values.}${}^{\scriptscriptstyle )}$ of a given integrable reduction of Einstein's field equations (e.g., the symmetry reduced vacuum Einstein equations or electrovacuum Einstein - Maxwell field equations). For a specific spectral problem\footnote{${}^{\scriptscriptstyle )}$ A construction of a class of possible associated linear systems with spectral parameters was suggested in \cite{Alekseev:1980a}${}^{,}$\cite{Alekseev:1980b}{}.
The simplest representative of these systems used in \cite{Alekseev:1985} differs only slightly from the vacuum linear system for generating function of Kinnersley and Chitre hierarchies of potentials constructed in \cite{KCIV:1978b} and generalized later for electrovacuum fields by Hauser and Ernst \cite{Hauser-Ernst:1979b}{}. An appropriate set of additional conditions providing equivalence of the entire spectral problem to the field equations was found in \cite{Alekseev:1985}{}.}${}^{\scriptscriptstyle )}$ used in this approach, the appropriately normalized  fundamental solution (which is equal to the unit matrix at some chosen initial or ``reference'' space-time point) possess in general four singular points on the spectral plane. These are algebraic branch points of the orders $1/2$ (two points) and  $-1/2$ (other two points). To select a holomorphic branch for any normalized fundamental solution, we always choose the cut on the spectral plane which consists of {\it two disconnected parts} (cuts) joining the pairs of branch points with opposite orders of branching  {}\footnote{${}^{\scriptscriptstyle )}$ This structure of the cut which consists of two disconnected parts is available in the most general case, for any solution, and this is the first important difference of this approach with the approach of Hauser and Ernst, who assumed that a fundamental solution possess only two singularities and therefore, the cut can consist of only one curve joining these two singularities. This ``minimization'' of the number of complex plane singularities can not be introduced in general and it obviously restricts the class of solutions and the domains where the solutions are defined (see one of the previous footnotes given to the discussion of the integral equation methods).\\[-2ex]}${}^{\scriptscriptstyle )}$.{}
On both parts of this compound cut, the holomorphic branch of the fundamental solution of our spectral problem possess some discontinuities (``jumps'') which can be described by two  coordinate-independent monodromy matrices (one matrix for each cut) whose components are functions of the spectral parameter only. These matrices determine the linear transformations of the fundamental solution after its analytical continuations along the paths each surrounding one of the endpoint singularities and joining the corresponding points on different edges of the cut. In the general case, the structures of these monodromy matrices are shown to be highly constrained (each matrix is equal to its inverse and it is a unit matrix plus a matrix of the rank equal to 1). Certain complete set of independent components of these matrices are called the monodromy data. For any vacuum solution the monodromy data consists of two functions of the spectral parameter: one of them should be holomorphic on one cut and another one - on the other, while for any electrovacuum solution we have two pairs of such functions.\footnote{${}^{\scriptscriptstyle )}$ If for stationary axisymmetric fields we impose the condition of regularity of the axis of symmetry, it can be shown that the  monodromy data functions in pairs should be analytical continuations of each other, i.e. we shall have only one function which characterize any vacuum solution and two functions which characterize any electrovacuum solution near the regular part of the axis of symmetry. These specific kind of data have been called as analytically matched ones. In these case these data can be simply related to the values of the Ernst potentials on this regular part of the symmetry axis and our description of the fields for such data can be related directly to the Hauser and Ernst approach (and therefore, with the Sibgatullin reduction also), where the regularity axis condition eventually occurred as an underlying assumption. However, in contrast to the axis data, the monodromy data functions can be determined also for solutions, which do not satisfy the regularity axis condition and therefore, which can not be considered in the Hauser-Ernst-Sibgatullin approach (e.g., the Levi-Civita and van Stockum solutions in stationary axisymmetric case, or various homogeneous or partially inhomogeneous singular cosmological solutions such as Kazner solution, or solutions for colliding plane waves such as, for example, Szekeres,  Khan and Penrose or Nutku and Halil solutions as well as many others). However, even in stationary axisymmetric case, the use of the monodromy data functions instead of the axis data allows to be not confined to the local region near the axis and to calculate (in principle, at least) the monodromy data functions from some boundary data given not necessarily on the axis.\\[-2ex]}${}^{\scriptscriptstyle )}$\\[0ex]

\noindent
This approach possess a number of important features, such as, in particular,
\begin{list}{$\diamond$}{\leftmargin4ex}
\item{
\emph{Generality:} this approach is available both for hyperbolic as well as for elliptic reductions of Einstein's field equations
and it does not need any additional ``working hypothesis'' besides the basic space-time symmetry ansatz\footnote{${}^{\scriptscriptstyle )}$This ansatz was described in the first footnote in this paper.}{}${}^{\scriptscriptstyle )}$.}
\item{\emph{The monodromy data as ``coordinates'' in the space of local solutions.} Similarly to the  scattering data in the inverse scattering transform, the monodromy data constitute a \emph{free} space of functional parameters  which  play the role of ``coordinates'' in the whole infinite-dimensional space of local solutions.}
\item{\emph{Direct and inverse problems of the monodromy transform.} For construction of the ``coordinate'' transformation in the space of solutions, we have to solve the ``direct'' and ``inverse'' problems which admit in general the unique solutions:
\begin{itemlist}
\item{\emph{Direct problem:} to find the monodromy data for any given solution. For solution of this problem it is necessary to find the fundamental solution of the associated linear system with a spectral parameter and with well defined initial conditions at a chosen reference point.\footnote{${}^{\scriptscriptstyle )}$Though a fundamental solution of the associated linear system always exists for any particular solution of the field equations, it is clear that this solution (and therefore, the monodromy data functions) can be found explicitly not for any choice of the solution, but for some specific cases only. That is why we say that given a solution, the monodromy data can be found at least, in principle. However, for many known cases this can be done explicitly.}${}^{\scriptscriptstyle )}${}}
\item{\emph{Inverse problem:} given monodromy data, to find the field components and potentials of the corresponding solution. The solution of this problem reduces  to solution of a decoupled system of linear singular integral equations of the Cauchy type on some compound cut on the spectral plane.\footnote{${}^{\scriptscriptstyle )}$The theory of such integral equations is well known. Using the standard method of analysis of such equations, it was shown that at least in some small enough region near the space-time reference point (where the fundamental solution and field components were normalized by some standard values), for any given monodromy data functions the solution of this system of linear singular integral equations always exists and is unique, what proves the existence and uniqueness of the solution of the inverse problem of the monodromy transform.}${}^{\scriptscriptstyle )}$ The coefficients of these integral equations possess simple algebraic expressions in terms of the monodromy data functions. In terms of  solution of these integral equations, all field components and potentials can be calculated in quadratures.}
\end{itemlist}}
\item{\emph{The linear algorithm for solution of initial and boundary value problems.} To determine the monodromy data functions, we do not need to know the corresponding solution  completely, but its  initial or boundary  data only. This allows (in principle) to reduce the solution of various nonlinear boundary value problems as well as  the Cauchy and characteristic initial value problems to the following three steps which represent pure linear problems \cite{Alekseev:1988}{}${}^{,}${}\cite{Alekseev:1993}{}:
\begin{romanlist}
\item{to find the monodromy data for the fundamental solution of  the linear ordinary differential equations along the boundary with the coefficients depending on the boundary data;}
\item{with this monodromy data, to find the solution of the basic system of linear singular integral equations solving the inverse problem of the monodromy transform;}
\item{to calculate the quadratures which determine the field components and potentials in terms of solution of these integral equations.}
\end{romanlist}}
\item{\emph{Infinite hierarchies of exact solutions:}
for large classes of monodromy data chosen as any rational functions of the spectral parameter, the inverse problem of the monodromy transform can be solved and an infinite hierarchies of solutions with any number of free parameters can be found explicitly (see  \cite{Alekseev:1988}{}${}^{,}${}\cite{Alekseev:1992}{} for   analytically matched rational monodromy data and\,\,\, \cite{Alekseev-Griffiths:2000a} for analytically not matched rational monodromy data).}
\item{\emph{Monodromy data and the solution generating methods:}
various soliton generating transformations as well as other known solution generating methods can be described conveniently in terms of the corresponding transformation of the monodromy data which can be given explicitly even in general form, i.e. without a preliminary specification of the background (seed) solution\cite{Alekseev:2001a}{}.}
\end{list}

\leftskip0ex\noindent
Various features of the monodromy transform approach mentioned above can give rise to its
diverse applications which examples can be found in
\cite{Alekseev:1992}${}^{-}$\!\!\cite{Alekseev-Belinski:2007c}{}.
\\[-3ex]

\section*{Later developments for vacuum and electrovacuum fields}
In the subsequent years, a number of more specific developments, new methods and interesting applications were found by different authors. In particular, these are \\[-2ex]

\subsubsection*{Algebro-geometrical methods and finite-gap solutions.}
For soliton solutions generated on some arbitrary chosen background, the fundamental solution of the associated spectral problem ``inherits'' the singularities corresponding to the background solution and acquires a number of simple poles corresponding to the number of solitons generated on this background.\footnote{${}^{\scriptscriptstyle )}$It seems useful to clarify here that there is no contradiction with what we said earlier that the fundamental solution of associated spectral problem in the most general case possess only four singular points (algebraic branch points) on the spectral plane. The point is that the fundamental solutions of associated spectral problem are defined up to a ``gauge'' transformation which is generated by multiplication of this  solution from the right by any  non-degenerate matrix function depending only on the (constant) spectral parameter. This matrix multiplier always can be chosen so that all other singularities (besides the four branch points mentioned above) will disappear. Such ``minimization'' of singularities takes place for the fundamental solution ``normalized'' by the condition that it takes the value of the unit matrix at some chosen space-time point (the reference point). In particular, for solitons the poles, which arise in the components of the fundamental solution of the spectral problem after application of a dressing method, can be removed by the normalizing multiplier. However, these poles do not disappear without any imprints: the poles corresponding to solitons just can be observed in the analytical structure of the monodromy data of the corresponding normalized fundamental solutions.}${}^{\scriptscriptstyle )}$ In contrast to the solitons associated with meromorphic (rational) functions on the (extended) spectral plane (Riemanian sphere), the distinguishing feature of the finite gap solutions is that the corresponding fundamental solutions of the associated spectral problem are meromorphic functions on some compact Riemanian surfaces of a given genus. These functions can be expressed in terms of the Riemann theta-functions and the whole construction becomes effective if the corresponding Riemann surface is of the hyperelliptic type because just in this case one can construct explicitly the basis of holomorphic differentials on this surface. For stationary axisymmetric vacuum Einstein equations such solutions were constructed in ``axiomatic'' manner by Korotkin and  Matveev\cite{Korotkin-Matveev:1989} and Korotkin\cite{Korotkin:1988, Korotkin:1993} using the methods of algebraic geometry developed earlier for other integrable systems in mathematical physics \cite{Novikov:1974}${}^{-}$\cite{Dubrovin:1981}{}.

\subsubsection*{Self-consistent solution for  a thin rigidly rotating dust disc.}
The finite-gap solutions had emerged also in the papers of Neugebauer and Meinel \cite{Neugebauer-Meinel:1993}${}^{-}$\!\!\ \cite{Neugebauer-Meinel:1995} in a quiet different way.  From the beginning, these authors aimed to solve a complicate but physically interesting problem of construction of the self-consistent solution for a rigidly rotating thin disc of dust in which both the ``internal'' and ``external'' parts of the solution (i.e. the density distribution of dust in the disk and the potentials of the external gravitational field) should be determined in a self-consistent way. In the mentioned above papers, this problem was reformulated as a Dirichlet boundary value problem for vacuum Ernst equation, and then it was shown that the solution of this problem can be reduced to the well known Jacobian inversion problem for ultraelliptic case, whose solution, as it is well known, can be expressed in terms of the corresponding Riemann theta functions. Many interesting properties of this rigidly rotating dust disk solution and other relativistic figures of equilibrium were described in the book of Meinel, Ansorg, Kleinwaschter, Neugebauer and Petroff\,\, \cite{Meinel-Ansorg-Kleinwaschter-Neugebauer-Petroff:2008} and in the references given there.

\subsubsection*{Colliding plane waves and solutions of  characteristic initial value problems.} At the very end of 80th, in a series of papers \cite{Hauser-Ernst:1989a}${}^{-}$\!\!\cite{Hauser-Ernst:1991}{}, Hauser and Ernst considered the characteristic initial value problem for colliding plane gravitational waves propagating initially through the Minkowski background. \footnote{${}^{\scriptscriptstyle )}$Many previous results and particular solutions for colliding plane gravitational waves in General Relativity with the corresponding references had been collected in the Griffiths book\cite{Griffiths:1991}}${}^{\scriptscriptstyle )}$ As a preliminary step, Hauser and Ernst presented in a nice form a general solution for colliding plane gravitational waves with collinear polarization (described by the linear Euler - Poisson - Darboux equation) expressed in terms of the integral Abel transform and formulated a new homogeneous Hilbert problem for this linear case.\footnote{${}^{\scriptscriptstyle )}$We recall here that the previous formulation of the homogeneous Hilbert problem suggested by these authors for stationary axisymmetric case can not be used for description of colliding plane waves on the Minkowski background because it
was based on the assumption of regularity of the axis of symmetry. However, in the case of
hyperbolic Ernst equations, for colliding plane waves the condition of regularity of the
symmetry axis transforms into the condition of an absence of focussing singularities for
waves, but as we know now, such singularities tipically appear as a result of interaction of
colliding plane waves on the Minkowski background.}${}^{\scriptscriptstyle )}$ After that, this homogeneous Hilbert
problem approach to the characteristic initial value problem for colliding gravitational plane
waves was generalized in these papers to the case of noncollinear polarization of colliding waves and the
solution of this problem was reduced to a matrix Fredholm equation of the second kind.\\[-1ex]

\leftskip0ex\noindent
On the other hand,  the monodromy transform approach  and the corresponding system of linear singular integral equations \cite{Alekseev:1985}${}^{,}$\cite{Alekseev:1988} (already discussed above) also admit a consideration of the solutions with focussing singularities which arise in the colliding plane wave solutions. Besides that, later some new form of linear singular integral equations was found \cite{Alekseev:2001b} and it occurred the most suited for solution of the characteristic initial value problems for the hyperbolic Ernst equations. These equations arose in further development of the monodromy transform approach where some dual representations for solutions of the associated spectral problem in terms of the ``scattering'' matrices were used. The condition of compatibility of these two representations gave rise to a new set of linear (quasi-Fredholm) integral equations equivalent to the dynamical parts of the symmetry reduced Einstein equations. Unlike the previously derived linear singular integral equations which kernels were expressed in terms of nonevolving monodromy data, the scalar kernels of new equations are constructed using the evolving (``dynamical'') monodromy data for the scattering matrices. These  integral equations were used in our papers with J.~Griffiths \cite{Alekseev-Griffiths:2001d}${}^{,}$\cite{Alekseev-Griffiths:2004} for construction of solutions of the characteristic initial value problem for plane gravitational as well as gravitational and electromagnetic waves colliding on the Minkowski background. In particular, in these papers it was shown that the dynamical monodromy data are completely determined by the parameters of the plane waves before their collision, and the corresponding solution of the integral equations allows to calculate the solution in the wave interaction region in quadratures.

\leftskip0ex

\section*{Integrability in the presence of matter fields}
Besides pure vacuum Einstein equations and electrovacuum Einstein - Maxwell equations, in the presence of some matter fields, the symmetry reduced Einstein's field equations also give rise to integrable systems. Below we mention these cases with some references and remarks.\\[-2ex]

\subsubsection*{Gravity coupled perfect fluid with a stiff matter equation of state $\varepsilon=p$.} In this case considered by Belinski \cite{Belinski:1979a}{}, the dynamical part of the field equations can be solved using Belinski and Zakharov inverse scattering approach. It is nice that these equations admit, in particular, the Friedman-Robertson-Walker solutions and this allows to analyze a picture of nonlinear wave dynamics on this cosmological background (the examples of such soliton ``cosmological''  waves also were considered in \cite{Belinski:1979a}). It is easy to observe, that the electromagnetic fields also can be included into this scheme and the corresponding symmetry reduced field equations for the Einstein - Maxwell fields and a neutral stiff matter fluid can be solved using the methods developed for solution of electrovacuum Einstein - Maxwell equations.\\[-2ex]

\subsubsection*{Einstein-Maxwell and dilaton fields.} The four-dimensional dynamical equations for gravitational, electromagnetic and dilaton fields can be expressed in the Kaluza and Klein form of 5-dimensional vacuum Einstein equations. The symmetry reduction of these equations for five-dimensional space-times with three commuting isometries take the same form of the matrix equations which were used by Belinski and Zakharov in their formulation of the inverse scattering method for vacuum fields, but for $3\times 3$- matrices. This was used by Belinski \cite{Belinski:1979b} and  Belinski and Ruffini \cite{Belinski-Ruffini:1980} for constructing the stationary axisymmetric solutions of the Einstein - Maxwell equations with a dilaton field applying the Belinski and Zakharov soliton generating technique.\\[-2ex]

\subsubsection*{The Einstein - Maxwell - Weyl equations for gravitational, electromagnetic and massless two-component spinor fields.}
For space-times with two commuting isometries, the symmetry reduced  Einstein-Maxwell-Weyl field equations also were found to be integrable \cite{Alekseev:1983}{}. In the reduced equations and in the corresponding spectral problem, the spinor field is represented by only one real ``harmonic'' function of two space-time coordinates. This function plays the role of a potential for certain pair of components\footnote{${}^{\scriptscriptstyle )}$These components correspond to the ``nonignorable'' coordinates which can be considered also as coordinates on the orbit space of the space-time isometry group.}${}^{\scriptscriptstyle )}$ of the spinor field current vector and it is a solution of the two-dimensional d'Alambert equation (for time-dependent fields, i.e. for hyperbolic reductions) or two-dimensional Laplace equation (for stationary fields, i.e. for the elliptic reductions). It is interesting to note, that an application of the dressing method leads to construction of solitons on arbitrary Einstein-Maxwell-Weyl background, however this soliton generating transformation does not change the mentioned above ``harmonic'' function which characterizes the spinor field and therefore, the solitons with spinor field can not be generated from pure vacuum or electrovacuum seed.

The Weyl spinor field enters very naturally into the constructions of the monodromy transform approach \cite{Alekseev:1985}{}, but it changes significantly the analytical structure of the fundamental solution of the  spectral problem\footnote{${}^{\scriptscriptstyle )}$In particular, for example, the singular points of this fundamental solution (which are algebraic branch points of the orders $\pm 1/2$ in the case of electrovacuum Einstein - Maxwell equations) may become the branch points of the infinite orders in the presence of Weyl spinor field.}${}^{\scriptscriptstyle )}$ and the corresponding linear singular integral equations  solving the inverse problem of the monodromy transform. Similarly to the electrovacuum Einstein - Maxwell case, these integral equations provide another way for constructing of solutions of Einstein - Maxwell - Weyl equations. For example,  these integral equations can be solved explicitly  for any choice of (analytically matched) as well as for some analytically not matched rational monodromy data{}\footnote{${}^{\scriptscriptstyle )}$The details of such calculations for electrovacuum case can be found in \cite{Alekseev:1988, Alekseev-Garcia:1996, Alekseev-Griffiths:2000a} and their generalization to the presence of the Weyl spinor field does not meet any principle difficulties.}${}^{\scriptscriptstyle )}$.

The integrability of the Einstein - Maxwell - Weyl equations for stationary axisymmetric fields was ``rediscovered'' later by Sibgatullin \cite{Sibgatullin:1983, Sibgatullin:1984b}, who generalized to the case of a presence of Weyl spinor field
the group-theoretic approach of Kinnersley and Chitre and the homogeneous Hilbert problem for Einstein - Maxwell equations of Hauser and Ernst. Following the method of Hauser and Ernst for the proof of the so called Geroch conjecture (i.e. the conjecture of transitive action of the group of internal symmetries in the space of solutions of the reduced Einstein - Maxwell equations), Sibgatullin also observed that these symmetry transformations do not change the potential for the Weyl spinor field and therefore, these symmetries act on the subspaces of the space of solutions labeled by this potential. He considered also the particular examples of stationary axisymmetric solutions of Einstein - Maxwell - Weyl equations and studied various singularities in these solutions.

\leftskip0ex\noindent
\section*{Integrability of the field equations in gravity, string gravity and supergravity models in four and higher dimensions}
In the literature of the last two decades, a lot of attention have been given to the studies of Einstein's gravity in $D > 4$ dimensions and to investigation of string gravity ($D=10$) and supergravity ($D=11$) models, their dimensional reductions and truncations. In many cases, the purpose of these studies was a construction of classical solutions describing the dynamics in the bosonic sector of these theories. For space-times with $D-2$ commuting isometries, some of these models were found to be integrable. However, not in all cases in which this integrability was already asserted, this gave rise to development of some effective solution generating methods and construction of various physically interesting solutions because some difficulties arise typically on this way. To clarify this, consider various integrable models separately.\\[-1ex]

\subsubsection*{Vacuum Einstein equations and electrovacuum Einstein - Maxwell equations in $D >4$ dimensions}

\leftskip0ex\noindent
The assumption of so large space-time symmetry ($D-2$ commuting Killing vector fields) occurs from physical point of view to be more restrictive than it was in $D=4$ case. In particular, this assumption excludes for $D>5$ the black hole solutions which need more than two variables for their description. Nonetheless, the symmetry reduced vacuum Einstein equations in $D=4$ and in higher dimensions represent the simplest case of the mentioned above integrable equations. The sigma-model-like form of these equations supplied with the conditions of reality and symmetry of the matrix of metric coefficients admits a direct application to this case of Belinski and Zakharov inverse scattering approach and corresponding soliton generating technique (or, may be, some alternative methods) without any significant changes of the procedures. However, even in $D=5$ case, the black hole solutions do not arise as solitons on the Minkowski background, as it was in $D=4$ case. In this case, the choice of appropriate set of coordinates and parameters and especially, of the appropriate seed (background) geometry becomes very important for construction of physically interesting solutions and this need more skill than in $D=4$ case. Nonetheless, in spite of these difficulties,  applications of such generalized methods to five-dimensional pure vacuum Einstein equations allowed to construct a number of interesting solutions as well as generalizations of the solutions constructed earlier by other methods. This concerns, first of all, a rich series of asymptotically flat field configurations of various compact black objects whose existence was discovered in space-times with $D=5$ dimensions and which were shown to be or found originally as soliton solutions of five-dimensional vacuum Einstein equations on some specially chosen (vacuum) background geometries. These are the black hole solutions which generalize the known Myers-Perry solution \cite{Myers-Perry:1986} and possess two angular momenta \cite{Pomeransky:2006}{}, various black ring solutions (the topology of the horizon is $S^1\times S^2$)  \cite{Emparan-Reall:2002}${}^{-}{}$\cite{Pomeransky-Sen'kov:2006}{},
black Saturn \cite{Elvang-Figueras:2007}{}, black di-rings \cite{Iguchi-Mishima:2007, Evslin-Krishnan:2007}{}, bicycling black ring \cite{Elvang-Rodriguez:2007}{} and others -- see also the surveys \cite{Emparan-Reall:2006, Emparan-Reall:2008} and  references there.

As it was shown in \cite{Yazadjiev:2006}{}, under a special ansatz for metric and electromagnetic potential, the symmetry reduced Einstein - Maxwell equations in $D=5$ space-times also can be presented in the form of sigma-model-like equations which can be solved using the Belinski and Zakharov spectral problem. In \cite{Yazadjiev:2006} the author presented also some interesting method which allows to construct the solutions with electromagnetic field from a pair of vacuum $D=5$ solutions, and illustrated this method by a construction of a dipole black ring starting from two vacuum black ring solutions.

\subsubsection*{Bosonic dynamics in the low-energy heterotic string theory and supergravity}

\leftskip0ex\noindent
The symmetry reduced dynamical equations which describe the bosonic sectors of some gravity models in four and higher dimensions also admit a construction of equivalent spectral problems based on the Belinski and Zakharov form of inverse scattering method. In particular, in the papers \cite{Bakas:1996a, Bakas:1996b} the Belinski and Zakharov soliton generating transformations were adapted for  solution of the equations for dilaton and axion fields coupled to gravity in $D=4$ space-times.

In more complicate cases, a part of the symmetry reduced dynamical equations also can take the sigma-model-like form. This immediately suggests the idea to use the Belinski and Zakharov linear system with a spectral parameter and to apply the inverse scattering approach for constructing solutions of these systems. Indeed, the existence of an associated linear system with a spectral parameter is one of the widely accepted evidences for integrability of a nonlinear system and this allowed to many authors to assert their discovery of integrability of the corresponding nonlinear field equations. However, the existence of such linear system itself does not lead directly to some effective methods for construction of  solutions because the integrability conditions of this linear system usually are not equivalent to the corresponding dynamical equations. Namely, the space of solutions of this linear system usually occurs larger than the space o solutions of the corresponding nonlinear equations, and some additional constraints (``reduction conditions'') should be imposed on the solutions of the linear system in order to provide an equivalence of a complete spectral problem (i.e. the linear system with the reduction conditions) to the original dynamical equations. However, for some cases in which a generalization of the Belinski and Zakharov linear system was found, it occurred that the solution of dynamical equations should possess some rather complicate (coset) structure for which the corresponding reduction conditions have not been found in an appropriate form and this does not allow to generalize the Belinski and Zakharov approach (or some other ones based on the similar ideas) to these cases. Just this situation evidenced, for example, in the case of electrovacuum Einstein - Maxwell fields in General Relativity. In this case, the Belinski-Zakharov-like form of the linear system with a spectral parameter for this case was constructed in \cite{Alekseev:1983a} (see also \cite{Alekseev:1988}{}), but the corresponding reduction conditions had not been found in a desirable form. The similar situation can arise in some other gravity models, such as, for example, in the case of the symmetry reduced minimal supergravity equations in $D=5$ dimensions. For these equations, the Belinski and Zakharov linear system was constructed in \cite{Figueras-Jamsin-Rocha-Virmani:2009}{}, where the authors observed, however, that after an application of the soliton generating transformations to a chosen seed, the transformed solution can leave the corresponding coset space (a discussion of this problem also can be found in \cite{Figueras-Jamsin-Rocha-Virmani:2009}).

The difficulties mentioned above had been solved in a series of cases considered in the framework of the monodromy transform approach (discussed in detail in one of the previous sections).
In this approach, we consider another type  of associated linear systems with a spectral parameter\,{} \cite{Alekseev:1980a,Alekseev:1980b,Alekseev:1985,Alekseev:1988}{}, for which the integrability conditions lead to the nonlinear dynamical equations in the self-dual form (instead of the sigma-model-like form). The advantage of the use of the linear systems of this type is that the corresponding reduction conditions (found in{}\cite{Alekseev:1985,Alekseev:1988}{}), which provide the equivalence of the complete spectral problem to the dynamical equations, take a simple algebraic form.  These supplementary conditions consist of two parts -- the conditions imposed on the structure of the canonical Jordan forms of matrix coefficients of the associated linear system and the condition of existence for this linear system of a matrix integral of certain structure. It is interesting to note that for all considered gravity models the structure of the linear system of this type and the corresponding reduction conditions possess the same form in which only the dimension of the linear system, the structure of the Jordan forms of the linear system and the symmetry properties of the first integral can be different for different models.

Just this ``self-dual'' form of the spectral problem was used by the author in the formulation of the inverse scattering approach and for construction of the soliton solutions for electrovacuum  Einstein - Maxwell equations in $D=4$ space-times \cite{Alekseev:1980a, Alekseev:1980b, Alekseev:1985, Alekseev:1988}. The same form of the spectral problem was constructed for the gravity models which can be reduced to the matrix analogues of the Ernst equations  for the cases in which the generalized (matrix) Ernst potential is (a) complex symmetric or (b) Hermitian matrix of arbitrary dimension\cite{Alekseev:2005a, Alekseev:2005b}.{}\footnote{${}^{\scriptscriptstyle )}$ It is known that some symmetry reduced string gravity models can be described by this kind of matrix equations.}${}^{\scriptscriptstyle )}$

More recently, an equivalent spectral problem of the same kind was constructed also for a complete system of symmetry reduced field equations which describes the bosonic dynamics in the low energy heterotic string effective theory in space-times of $D$ dimensions with $D-2$ commuting isometries\cite{Alekseev:2009}. This system, besides the gravitational field,  includes also the dilaton, antisymmetric gauge field and arbitrary number of Abelian gauge vector fields. The spectral problem found there allows to construct for these equations some types of soliton solutions, to generalize for this case the monodromy transform approach and the integral equations methods developed earlier for constructing solutions of symmetry reduced vacuum Einstein equations and Einstein - Maxwell equations in General Relativity in four dimensions.

\section*{Concluding remarks}
A brief outline of thirty years of studies of integrable reductions of Einstein's field equations including integrability of symmetry reduced gravity, string gravity and supergravity models in four and higher dimensions  given above clearly can not be an exhaustive and systematical survey. Inevitably, this can consist of some fragments of the history only. The emphasis here was made on the origins of different approaches and  comparison of providing opportunities as well as on some papers and results more or less ``forgotten'' or misinterpreted in other surveys, numerous introductions and discussions. On the other hand, it is worth to mention that some of the results which were described in other surveys very carefully, have not been mentioned here with the same amount of details. Hopefully therefore, the present survey can represent a useful supplement to the already existing ones.

It necessary to note also that the power of various mathematical methods mentioned or discussed in this paper had not been exhausted by the already known applications. And, without any doubts, we can expect further developments of these methods and their new interesting applications in General Relativity as well as in the low-energy effective string gravity and supergravity models in four and higher dimensions.

\section*{Acknowledgements}
The author is thankful to the Organizing Committee of
MG12 for partial financial support for participation in the Meeting. This work was supported in parts by the Russian Foundation for Basic Research (grants 08-01-00501, 08-01-00618, 09-01-92433-CE) and the program ``Mathematical Methods of Nonlinear Dynamics'' of the Russian Academy of Sciences.

\end{document}